\documentstyle[aps,prl]{revtex} 
\draft 
 
\def\lsim{\mathrel{\raise.3ex\hbox{$<$\kern-.75em\lower1ex\hbox{$\sim$}}}} 
\def\gsim{\mathrel{\raise.3ex\hbox{$>$\kern-.75em\lower1ex\hbox{$\sim$}}}}

\newcommand{\be}{\begin{equation}}
\newcommand{\ee}{\end{equation}}
\newcommand{\bea}{\begin{eqnarray}}
\newcommand{\eea}{\end{eqnarray}}

\begin{document} 
 
\twocolumn[\hsize\textwidth\columnwidth\hsize\csname
@twocolumnfalse\endcsname

\title{Possible Evidence For Axino Dark Matter In The Galactic Bulge} 
\author{Dan Hooper$^1$ and Lian-Tao Wang$^2$} 
\address{
$^1$Astrophysics Department, University of Oxford, Oxford, UK;
$^2$Department of Physics, University of Wisconsin, Madison, USA.}
\date{\today} 
 
\maketitle 
 
\begin{abstract}

Recently, the SPI spectrometer on the INTEGRAL satellite observed
strong 511 keV line emission from the galactic bulge. Although the
angular distribution (spherically symmetric with width of $\sim
9^{\circ}$) of this emission is difficult to account for with
traditional astrophysical scenarios, light dark matter particles could
account for the observation. In this letter, we consider the
possibility that decaying axinos in an R-parity violating model of
supersymmetry may be the source of this emission. We find that $\sim
1-300\, \rm{MeV}$ axinos with R-parity violating couplings can
naturally produce the observed emission. 

\end{abstract}

]

\section{Introduction}

The SPI spectrometer on the INTEGRAL (INTErnational Gamma-Ray Astrophysics
Laboratory) satellite has made the observation of a bright
($9.9^{+4.7}_{-2.1} \times
10^{-4}\,\rm{ph}\,\rm{cm}^{-2}\,\rm{s}^{-1}$), 511 keV gamma-ray
emission line from the galactic bulge \cite{511}. The emission is
consistent with being spherically symmetric, with a
full-width-half-maximum of about $9^{\circ}$($6^{\circ}-18^{\circ}$ at
2$\sigma$ confidence). The 3 keV width of the line is dominated by $e+
e^-$ annihilations via positronium formation \cite{milne}. These
findings are in agreement with earlier observations, such as those by
the OSSE experiment \cite{previous}. 

This observation of bright 511 keV emission from the galactic bulge
has been quite difficult to explain with traditional
astrophysics. Most potential sources considered do not produce a
sufficient number of positrons (such as neutron stars, black holes
\cite{compact}, radioactive nuclei from supernovae, novae, red giants
or Wolf-Rayet stars \cite{stars}, cosmic ray interactions with the
interstellar medium \cite{ism}, pulsars \cite{pulsars} or stellar
flares) and those which may possibly be capable of producing the
required number (type Ia supernovae \cite{debate,casse} or hypernovae
\cite{sn2003,casse}), may not be capable of filling the entire
galactic bulge \cite{starbursts}. In particular, the rate at high
altitude is likely to be too low to explain the observed extension of
the 511 keV source \cite{pohl}.  

Given this difficulty, alternative explanations should be
considered. In Ref.~\cite{boehm}, it was suggested that light (1-100
MeV) dark matter particles annihilating in the galactic bulge could
explain the observed emission. In this letter we, instead, consider
light {\it decaying} dark matter particles. In particular, we consider
the supersymmetric partner of the axion\cite{axion}, the axino, in
R-parity violating supersymmetric models. 


In the Minimal Supersymmetric Standard Model (MSSM), a $Z_2$ symmetry, R-parity\cite{rparity}, is usually imposed to
forbid dimension four 
operators which lead to fast proton decay \cite{protondecay}. A by-product of an exact R-parity is that the Lightest Supersymmetric
Particle (LSP) is stable and often a good candidate for cold dark
matter. Although R-parity is an elegant way of suppressing proton
decay, it does not have to be the only way that nature can choose to
do so. In particular, baryon parity is sufficient to make the proton stable. In this case, the coupling strengths of other
R-parity violating operators, such as $\lambda_{ijk}L_i L_j E^c_k$,
are  much less constrained. For a summary of R-parity violating
coupling constraints, see Ref.~\cite{rpv}. 


A typical dark matter candidate provided by R-parity conserving supersymmetry is a neutralino LSP. Neutralinos are the
superpartners of the neutral gauge bosons and Higgs bosons and have masses constrained by direct searches to be greater than $\sim 30$ GeV, too heavy to produce a large flux of thermal positrons. Of course, with sizable R-parity violation, the neutralino will have a short lifetime and cease to be a good candidate for
cold dark matter.

With the Peccei-Quinn (PQ) mechanism \cite{axion} providing a natural solution to the
strong-CP problem in a supersymmetric theory, the existence of an
axino is inevitable. The axino's mass is
expected to be considerably lower than the electroweak scale, perhaps
in the keV to several GeV range and is capable of providing the observed quantity of dark matter given a low reheating temperature \cite{axinodark,kim01,axinodecay,covi99,covi01}. Since the axino is in the
same supermultiplet as the axion, its couplings to matter fields
are generically suppressed by $f_a^{-1}$, where $f_a$ is the PQ
symmetry breaking scale $\sim 10^9 - 10^{12}$ GeV. As we shall see in
detail in this paper, due to this large suppression, axinos could be
considered {\it stable} during the history of the universe even in
the presence of sizable R-parity violating couplings. Furthermore, a long-lived MeV-GeV axino, such as we consider in this letter, would be heavy enough to constitute a good candidate
for cold dark matter \cite{covi99,covi01,axinoreheat}.

A possible concern for axinos in the early universe is their effect on the light element abundances \cite{lightelementscorrect,reheating,lightelements}. With a long-lived axino, however, axino decays do not threaten these
observations. Furthermore, with the introduction of R-parity violation
in our scenario, we allow for relatively fast Next-to-Lightest
Supersymmetric Particle (NLSP) decays into Standard Model
particles. Thus the epoch of SUSY particle decays is over prior to
nucleosynthesis and the light element abundances remain unaffected
\cite{lightelementsrviolation}. 

Before going into a more detailed study of decaying axinos, we briefly
consider here the 
possibility of decaying gravitinos. Gravitinos also have highly
suppressed couplings to matter ($M_P^{-1}$ in the kinematical regime of
interest) and could be a good candidate for cold dark matter.  We have
found, however, that for the case of trilinear 
R-parity violating terms, the gravitino lifetime is too long to account for the
observed 511 keV emission (see section II). In particular, 
the gravitino lifetime is estimated to be $\tau_{3/2} \sim
10^{31}\,\mbox{sec}\, \left(m_{\tilde{l}}/100\,\mbox{GeV} \right)^4 \,
\left(0.1\,\mbox{GeV}/m_{3/2}\right)^7 \, \left(0.1/\lambda\right)^2$.

 

\section{Decaying Dark Matter}
If light dark matter particles constitute the galactic halo, decays of
such particles can produce positrons which eventually annihilate
producing the observed 511 keV emission. In this section, we calculate
the lifetime of a light decaying dark matter particle needed to
account for the observed 511 keV flux. 

We note that although the observed emission line is energetically
narrow, indicative of positrons annihilating at rest, positrons
resulting from dark matter decays need not be at rest, as they will
stop easily via ionization losses before annihilating as long as their
initial energy is less than $\sim100\,\rm{MeV}$ \cite{boehm}. This
energy corresponds to decaying dark matter particles of a few hundred
MeV or less. 

The dark matter distribution in the galactic halo is traditionally
parameterized by 
\begin{equation}
\rho(r) \propto \frac{1}{(r/a)^{\gamma}
  [1+(r/a)^{\gamma}]^{(\beta-\gamma)/\alpha}},  
\end{equation}
where $\alpha$, $\beta$ and $\gamma$ are given by the choice of halo
profile and $a$ is the distance from the galactic center at which the
inner power law breaks. In the galactic bulge, $r \ll a$, and the
parameterization reduces to $\rho(r) \propto
\frac{1}{(r/a)^{\gamma}}$. Integrating over the line-of-sight of the
observation, and averaging over the angular resolution of SPI ($\sim
2^{\circ}$), the angular distribution of 511 keV gamma-rays from dark
matter decays for a given halo profile can be compared to the
observations of SPI/INTEGRAL. In figure 1, we show these results. We
find that for a cusped profile, $\gamma \simeq 1.2$, the data is well
fitted. This is in agreement with the results of recent high resolution N-body simulations~\cite{nbody}. Note that 
in Ref.~\cite{boehm}, the best fit was found for $\gamma \simeq
0.6$. This is because for annihilating dark matter, the annihilation
rate is proportional to the density squared, rather than simply the
density. 


\begin{figure}[thb]
\vbox{\kern2.4in\includegraphics{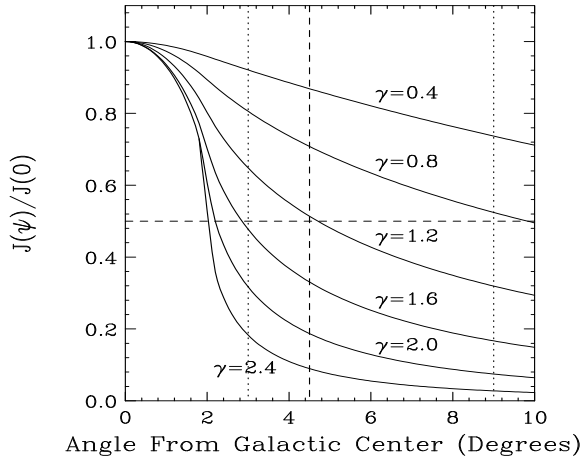}}

\caption{The angular distribution of 511 keV $\gamma$-rays from
  decaying dark matter averaged over the $2^{\circ}$ angular
  resolution of the SPI spectrometer on INTEGRAL for several halo
  profiles. SPI's observation indicates a full width half maximum of
  $9^{\circ}$ with a $6^{\circ}-18^{\circ}$ 2$\sigma$ confidence
  interval. Shown as vertical dashed and dotted lines are the central
  value and 2$\sigma$ limits of the angular widths found by SPI. To
  agree with this data, a cusped halo model with $\gamma \sim$ 0.8-1.5 is
  favored.} 
\end{figure}

Normalizing the halo profile to the local dark matter density, we obtain
\begin{equation}
\rho(r) \simeq  \frac{0.3\, \rm{M_{\odot}}/\rm{pc}^3}{(r/1\, \rm{kpc})^{\gamma}}, 
\end{equation}
where $\gamma \simeq1.2$. The total mass within the $9^{\circ}$ circle
observed by INTEGRAL is then 
\begin{equation}
M = \int^{670\, \rm{pc}}_{0} \rho(r)  4 \pi r^2 dr  \simeq 1 \times
10^9 \, \rm{M}_{\odot} \cong 1.5 \times 10^{66} \, \rm{GeV}. 
\end{equation} 
The rate at which dark matter decays in this region is simply the
total mass divided by the mass and lifetime of the dark matter
particle. Also, the rate of decays can be matched to the observed flux
of 511 keV gamma-rays. Comparing these two quantities yields 
\begin{equation}
\frac{1.5 \times 10^{66} \, \rm{GeV}}{m_{\rm{dm}} \tau_{\rm{dm}}} \sim
\frac{1}{2} \Phi_{\gamma, 511} 4 \pi R_{\rm{GC}}^2. 
\end{equation}
Here, $\Phi_{\gamma, 511}$ is the observed 511 keV gamma-ray flux and
$R_{\rm{GC}}$ is the distance of the Earth from the galactic
center. Inserting the observed flux ($\Phi_{\gamma, 511}\simeq 10^{-3}
\, \rm{ph}/\rm{cm}^2\rm{s}$) and $R_{\rm{GC}} \simeq 2.5 \times
10^{22}\,\rm{cm}$, we arrive at 
\begin{equation}
\label{datalifetime}
\tau_{\rm{dm}} \sim 4 \times 10^{26}\, \rm{seconds} / m_{\rm{dm}}(MeV),
\end{equation}
which is considerably longer than the age of the universe.

\section{Axino Decay}

Via R-parity violating couplings, axino LSPs may decay to Standard
Model particles. The decay width for such processes depends on the
axion model and the nature of the R-parity violation considered. 

There are two classes of invisible axion models, KSVZ\cite{ksvz} and
DFSZ\cite{dfsz}. Both of these introduce at least one extra field,
$\Phi$, which breaks the Peccei-Quinn (PQ) symmetry at a high 
scale, $f_a$. Supersymmetric versions of these models give rise to two
corresponding classes of axinos.  

In the first of these scenarios (KSVZ), the field $\Phi$ couples to a
pair of heavy quark states via the superpotential 
coupling, $W=\lambda \Phi Q \bar{Q}$. 
In the second scenario (DFSZ), Standard Model fermions carry PQ
charge. However, they do  
not have direct couplings to the PQ field. The PQ field, however,
couples to the Higgs sector which contains two Higgs 
doublets. In both scenarios, the axino has a
gaugino-gauge boson coupling proportional 
to the PQ anomaly \cite{covi99}
\begin{eqnarray}
{\mathcal{L}}_{\tilde{a} \lambda A} &=& i \frac{\alpha_Y C_{aYY}}{16 \pi
(f_a/N)} \bar{\tilde{a}} \gamma_5 [\gamma^{\mu}, \gamma^{\nu}]
\tilde{B} B_{\mu \nu} \nonumber \\
&+& i \frac{\alpha_s }{16 \pi
(f_a/N)} \bar{\tilde{a}} \gamma_5 [\gamma^{\mu}, \gamma^{\nu}]
\tilde{g}^b G_{\mu \nu}^b.
\end{eqnarray}
$C_{aYY}$ is a model dependent number of ${\mathcal{O}}(1)$.  

We will be mainly interested in axino-fermion-sfermion couplings, in particular 
axino-lepton-slepton couplings, $g_{\tilde{a} l \tilde{l}}$, in both of
those scenarios.  In the KSVZ scenario, there is no direct coupling
between the axino and Standard Model fields since they 
are not charged under $U(1)_{PQ}$. However, such an coupling could be
induced at loop-level via a Bino-$A_{Y}$-lepton loop
\cite{covi01} as 
\be
\label{gksvz}
|g_{\tilde{a} l \tilde{l}}| \propto \frac{\alpha_Y^2 C_{aYY}^2}{
  \pi^2} \frac{M_{1}}{f_a} \log \left( \frac{f_a}{M_1}\right).
\ee
On the other hand, in the DFSZ scenario, since the axino has a
higgsino component, there is going to be a direct coupling
\be
\label{gdfsz}
|g_{\tilde{a} l \tilde{l}}| \sim g\frac{v}{f_a}, 
\ee
where $g$ is of the size of the
gauge coupling suppressed by the mixing angle between higgsinos and
gauginos. 
For our estimate, it is useful to pull out the common factor in $g_{\tilde{a} l
  \tilde{l}}$ (with the 
rough order of magnitude identification $v \sim M_1$ in mind)
and write $g_{\tilde{a} l \tilde{l}} = \hat{g} v/f_a $. From
  Eq.~\ref{gksvz} and Eq.~\ref{gdfsz}, we can roughly
estimate $\hat{g}\sim  10^{-2}$ in 
the DFSZ case and $\hat{g} \sim 10^{-4}$ in the KSVZ case.

The exact decay width depends on the details of the model, such as
the Higgs potential and the spectrum and mixings of the
superpartners. Rather than going into a detailed study, we give here
an order-of-magnitude estimate. The axino life time is estimated to be 
\begin{eqnarray}
\tau_{\tilde{a}} &\sim& 10^{20} \ \mbox{sec} \times \left(
\frac{10 \,\mbox{MeV}}{m_{\tilde{a}}} \right)^5  \nonumber \\
&\times& \left(
\frac{m_{\tilde{l}}}{\mbox{TeV}}\right)^4 \left( \frac{f_a}{10^{11}
  \,\mbox{GeV}} \right)^2 \left( \frac{0.1}{\lambda}\right)^2
\hat{g}^{-2}. 
\end{eqnarray}
$\lambda$ is the R-parity violating leptonic trilinear coupling which
appears in the superpotential as $W = \lambda_{ijk} L_i
L_j E_k^c$. Positrons could be produced in the decays $\tilde{a}
\rightarrow \nu_{\tau} e^+ e^-$ or $\tilde{a}\rightarrow \nu_{\mu} e^+
e^-$ which result from the couplings $\lambda_{311}$ and
$\lambda_{211}$, respectively. $\lambda_{211}$ is constrained by
charge current universality as $\lambda_{211} \lsim 0.1
(m_{\tilde{e}_R}/200 \, \rm{GeV})$. $\lambda_{311}$ is constrained by
$\Gamma({\tau \rightarrow e \nu \bar{\nu}}) / \Gamma({\tau \rightarrow \mu
  \nu \bar{\nu}})   $
as $\lambda_{311} \lsim 0.12 (m_{\tilde{e}_R}/200 \,
\rm{GeV})$. See Ref.~\cite{rpv} for details. For axinos with a mass in
the range of 1-300 MeV, 
for a wide range of couplings
it is possible
to obtain the desired lifetime found in Eq.~\ref{datalifetime}.  

In addition to trilinear R-parity violating decays, bilinear 
couplings of the form $\mu_i L_i H_2$ could also contribute to the 511
keV gamma-ray production. The impact of this term on the axino
lifetime was studied in Ref.~\cite{kim01}. 
This bilinear term induces mixing between higgsinos (and hence
photinos) and neutrinos. Therefore, axinos can decay via
$\tilde{a}\rightarrow \gamma + \nu$. $\mu_i$ is constrained by the
diffuse gamma ray background to be $<$ keV \cite{kim01}. In this case, the
axino lifetime is about $10^{25}$ sec which is potentially interesting.
However, since this decay produces $\gamma + \nu$ rather than
positrons, this mode is useful for explaining the 511 keV line
structure only if 
the mass of the axino is precisely 1022 keV. Such a scenario is indeed
fine-tuned, although it is technically natural.    
Another potentially important mode induced by the bilinear coupling is
$\tilde{a} \rightarrow \tau^+ + \pi^{-}$. However, it is kinematically
forbidden for the range of axino masses we are interested in.

\section{Discussion and Conclusions}

In this letter, we have demonstrated that a light axino (1-300 MeV), in
either the KSVZ or DFSZ axion models,
with trilinear R-parity violating couplings could be responsible for the 511 keV line
emission observed from the galactic bulge. In this scenario, axinos
constitute the major component of the cold dark matter and are
present in the galactic halo with a cusped distribution ($\gamma
\sim 1.2$).  

At this time, we can not exclude the possibility that poorly
understood conventional astrophysics is responsible for the observed
positron production in the galactic bulge. To differentiate such a
scenario from more exotic sources, such as light decaying particles,
future tests must be made \cite{dwarf}. Additionally, if decaying
axinos are the source of the 511 keV emission, signatures of
supersymmetry, and R-parity violation will likely be
observed at the LHC. 

As this letter was being completed, an article appeared which also discussed decaying particles as the source of the observed 511 keV emission \cite{new511}. They considered decaying sterile neutrinos with rather constrained mixing parameters. They also discussed decaying scalars with gravitational strength interactions. 

{\it Acknowledgments}: DH is supported by the Leverhulme Trust. LW is
supported in part by U.S. Department of Energy under grant
DE-FG02-95ER40896. We would like to thank D. Chung and L. Roszkowski for useful
discussions.

\end{document}